\documentclass{article} % For LaTeX2e
\usepackage{iclr2026_conference,times}
\usepackage{amsmath,amsfonts,bm}

\def\eqref#1{equation~\ref{#1}}
\def\1{\bm{1}}

\DeclareMathAlphabet{\mathsfit}{\encodingdefault}{\sfdefault}{m}{sl}
\SetMathAlphabet{\mathsfit}{bold}{\encodingdefault}{\sfdefault}{bx}{n}

\newcommand{\E}{\mathbb{E}}

\usepackage{hyperref}
\usepackage{url}
\usepackage{graphicx}
\usepackage{float}
\usepackage{xr-hyper}
\title{CERBERUS: A Three-Headed Decoder \\for Vertical Cloud Profiles}

\author{Emily K. de~Jong, Nipun Gunawardena, Kevin Smalley, Hassan Beydoun, \& Peter Caldwell \\
Lawrence Livermore National Laboratory\\
Livermore, CA 94550 USA\\
\texttt{\{dejong5, gunawardena1, smalley5, beydoun1, caldwell19\}@llnl.gov}\\
}
\iclrfinalcopy % Uncomment for camera-ready version, but NOT for submission.
\begin{document}

\maketitle

\begin{abstract}
Atmospheric clouds exhibit complex three-dimensional structure and microphysical details that are poorly constrained by the predominantly two-dimensional satellite observations available at global scales. This mismatch complicates data-driven learning and evaluation of cloud processes in weather and climate models, contributing to ongoing uncertainty in atmospheric physics. We introduce CERBERUS, a probabilistic inference framework for generating vertical radar reflectivity profiles from geostationary satellite brightness temperatures, near-surface meteorological variables, and temporal context. CERBERUS employs a three-headed encoder–decoder architecture to predict a zero-inflated (ZI) vertically-resolved distribution of radar reflectivity. Trained and evaluated using ground-based Ka-band radar observations at the ARM Southern Great Plains site, CERBERUS recovers coherent structures across cloud regimes, generalizes to withheld test periods, and provides uncertainty estimates that reflect physical ambiguity, particularly in multilayer and dynamically complex clouds. These results demonstrate the value of distribution-based learning targets for bridging observational scales, introducing a path toward model-relevant synthetic observations of clouds.
\end{abstract}

\section{Introduction}
Atmospheric cloud processes occur at scales that are poorly resolved by a majority of observations available for model learning and validation. The satellite record is extensive in spatiotemporal coverage but primarily consists of 2D top-of-atmosphere perspectives. Meanwhile, vertically-resolved measurements of cloud properties are confined to sparse ground-based sites and \textit{in situ} aircraft measurements \citep{lamb_perspectives_2026}. This scale mismatch is one reason why clouds continue to drive uncertainty in both weather and climate predictions \citep{boucher_clouds_2013, morrison_confronting_2020}.

Recent work has leveraged the polar-orbiting radar CloudSat as a target for conditionally-generated cloud structures using GANs \citep{leinonen_reconstruction_2019}, U-Nets \citep{bruning_artificial_2024}, and masked-autoencoders (MAEs) \citep{girtsou_3d_2025, ermis_global_2025}. However, currently these approaches target only daytime clouds and may be deterministic, leaving ground-based measurements unexploited and uncertainty unquantified. This work introduces CERBERUS (\textbf{C}loud \textbf{E}stimation with vertically-\textbf{R}esolved \textbf{Be}ta-distributed \textbf{R}etrievals of \textbf{U}ncertainty and \textbf{S}tructure), a probabilistic data-driven framework for inferring vertical radar reflectivity conditioned on both space-based imagery and near-surface meteorological context. CERBERUS uses three prediction heads to output the probability of reflective cloud and two parameters of the reflectivity distribution at each altitude.  %By explicitly modeling the zero-inflated nature of cloud reflectivity, CERBERUS predicts vertically-resolved distributions that reflect the ill-posedness of 2D-to-3D inference as well as cloud regime complexity. 
This work illustrates CERBERUS at the Atmospheric Radiation Measurement (ARM) site in Oklahoma, USA, demonstrating scalable estimation and uncertainty quantification of cloud vertical profiles that can facilitate atmospheric model evaluation and calibration.

\section{Data \& Methods}
\subsection{Dataset \& Preprocessing}
\textbf{Target data: } Reflectivities and cloud-top products from the ARM Ka-band zenith-pointing radar (KAZR) at the Southern Great Plains (SGP) site are collected from January 2020--March 2025 \citep{KAZR_SGP, kollias_millimeter-wavelength_2007}. Data are quality-controlled based on signal-to-noise, resampled to 5-minute averages, and interpolated to 128 equispaced altitudes from 160~m to 15~km and smoothed with a Savitzky-Golay filter \citep{savitzky_smoothing_1964} with order 3 and window length 50.

\textbf{Input data: } Inputs to the inference model include 30-minutely 2D brightness temperatures from the GOES-16 satellite at 13.3~$\mu$m, 11.9~$\mu$m (SW), 11.2~$\mu$m (IR), 8.4~$\mu$m, 6.8~$\mu$m, and 3.9~$\mu$m (SIR), plus the visible reflectance (0.65~$\mu$m) \citep{GOES_SGP}. These seven fields are remapped to an $8\times 8$ grid at $0.02^\circ$-resolution centered at the ARM SGP site. Cloud-top-height (CTH) is retrieved using the VISST algorithm from the same datastream to confirm consistency of observed clouds between the radiometer (GOES) and radar (KAZR) measurements. In addition we consider five near-surface meteorological variables from hourly MERRA-2 reanalysis \citep{MERRA2} sampled at the SGP KAZR site: 10~m temperature (T), winds (u, v), and relative humidity (rh), as well as surface pressure (P0). Time-of-day and day-of-year are sine-encoded to account for seasonality and diurnal cycle, but no positional encoding is included as this experiment targets a single location.

\textbf{Data selection: } To ensure consistency and detectability in the observed clouds between the radar reflectivity target and the satellite fields, we impose 4 filters on the 30~min paired KAZR-GOES retrievals used for training and evaluation (see section~\ref{sec:Data}, Figure~\ref{fig:confusion}). We utilize an 80/20 training/validation random data split over 2020--2024 data (18,000 target-profile pairs), reserving the remaining 3 months of data from winter 2025 for testing (1600 pairs).

\textbf{Normalization: } 
All data utilize a min-max normalization. Non-cloudy values of the normalized reflectivity target dataset are zero-filled, corresponding to a detection threshold of -60~dBZ. This leads to a zero-inflated (ZI) target distribution with all values between 0 and 1 inclusive (Figure~\ref{fig:pdf}), motivating the choice of a zero-inflated beta distribution (ZIB) to model these data.

\subsection{The CERBERUS Model Structure \& Training}
\begin{figure}[!t]
    \centering
    \includegraphics[width=\linewidth]{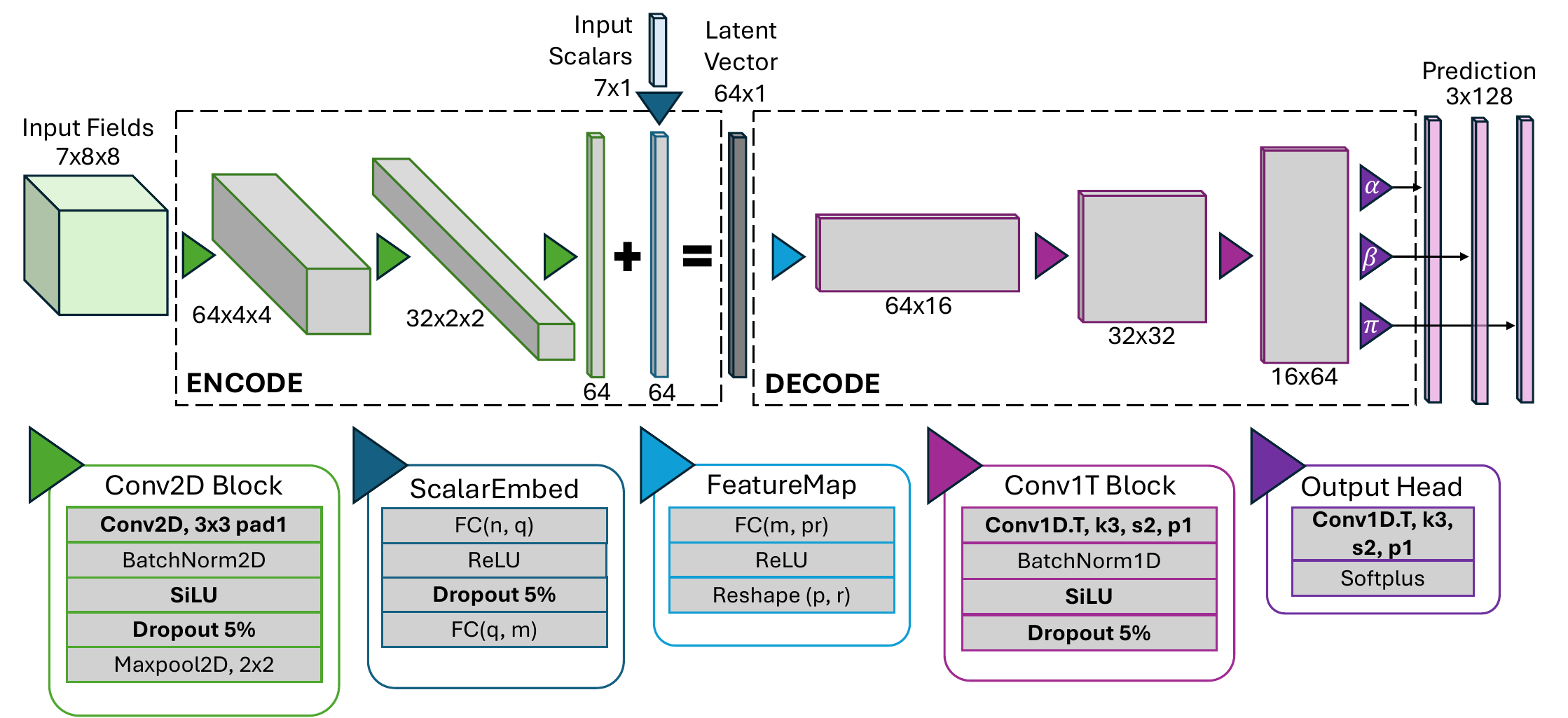}
    \caption{Illustration of the CERBERUS three-headed architecture and neural network operations. Bold text indicates hyperparameters that were optimized.}
    \label{fig:diagram}
\end{figure}
The structure of the 3-headed CERBERUS encoder-decoder architecture is illustrated in Figure~\ref{fig:diagram}. The brightness temperature fields are encoded via convolutions and then summed with embedded near-surface scalars in a FiLM-like approach \citep{perez_film_2017}. The decoder then projects, reshapes, and transforms the resulting latent vector, with the final layer utilizing three separate convolutional output heads to predict the three parameters of a zero-inflated beta (ZIB) reflectivity distribution at each of the 128 target altitudes. Hyperparameters including dropout, CNN activation function, and kernel size were optimized \citep{akiba_optuna_2019}. 

CERBERUS has similarities to related UNet and SatMAE models for reconstructing 3D clouds \citep{girtsou_3d_2025, ermis_global_2025, bruning_artificial_2024}, such as mapping 2D satellite fields to a latent space. However, the three-headed decoder of CERBERUS allows for simultaneous learning of uncertainty in the reflectivity profiles, unlike prior deterministic approaches. A comparison against deterministic and non-ZI baselines (section~\ref{sec:baselines}, Figures~\ref{fig:crps} and \ref{fig:baselines}) supports the added value of this three-headed probabilistic structure. While previous models used RMSE as their training objective, this probabilistic approach uses the negative log-likelihood of the observation $y$ in the predicted distribution:
\begin{align}
    \mathcal{L}(y, (\alpha, \beta, \pi )) = \begin{cases}
        -\log(\pi), &  y =0\\
        -\log\big(1-\pi - \mathcal{B}[\alpha,\beta](y)\big), & y>0
    \end{cases}
\end{align}
where $\pi$ is the predicted probability of non-cloudiness and $\mathcal{B}[\alpha,\beta]$ is the beta distribution with predicted parameters $\alpha, \beta$. We add a small scalar $\epsilon=10^{-3}$ to all $\log$-arguments for stability. Model weights are trained using the Adam optimizer with batch size 100 and initial learning rate $10^{-3}$ for a maximum of 50 epochs, selecting the model with the smallest validation loss for evaluation (Figure~\ref{fig:loss}).

\section{Results}
CERBERUS demonstrates robust performance across both conditional classification of cloudy altitudes (ROC-AUC=0.957) and regression ($R^2 > 0.6$; Figure~\ref{fig:rsq}). IR and near-IR brightness temperatures contribute most to model accuracy, with the visible reflectance (only available during daytime measurements) being the least important GOES field (Figure~\ref{fig:pfi}). Out of the scalar conditions, the 10~m temperature contributes most to model performance and is indicative of thermodynamic and boundary layer characteristics. Near-surface winds and humidity, by contrast, may be redundant with information already embedded in the IR observations. 

\begin{figure}[!t]
    \centering
    \includegraphics[width=0.4\linewidth]{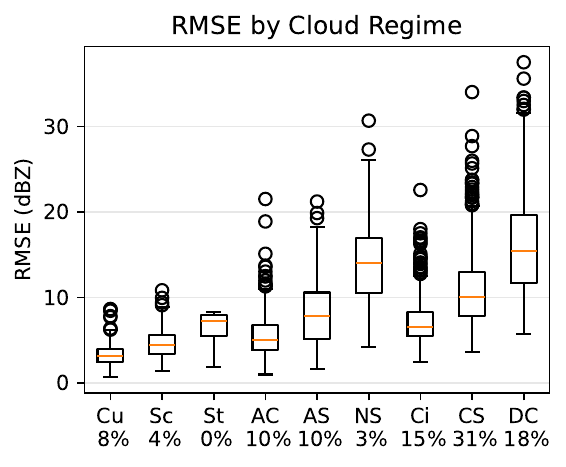}
    \caption{Statistics of per-sample RMSE (dBZ) for each cloud regime (see Table~\ref{tab:rmse_table}) over the validation set: median (orange), interquartile-range ($IQR$; box), $Q_1 / Q_3 \pm 1.5 IQR$ (whiskers), and outliers (circles). Regime abbreviations defined in Table~\ref{tab:rmse_table}.}
    \label{fig:rmse_by_regime}
\end{figure}

\begin{figure}[!t]
    \centering
    \includegraphics[width=0.8\linewidth]{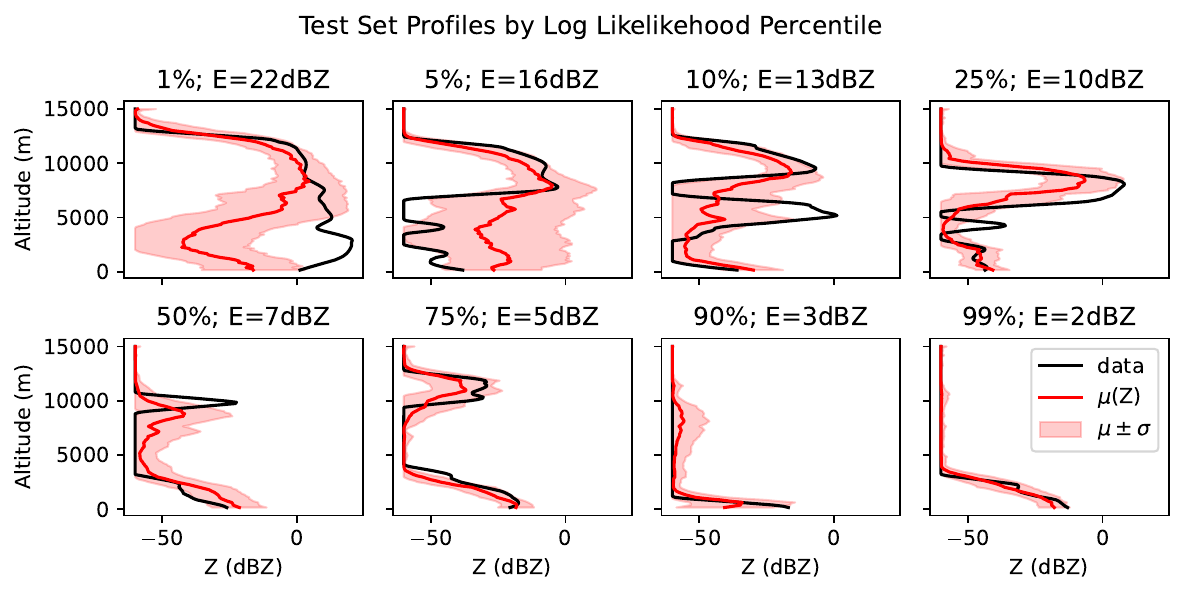}
    \caption{Test data and predicted reflectivity mean and uncertainty as a function of altitude, ranked according to quantile using altitude-scale energy (analogous to RMSE; see section~\ref{sec:metrics}).}
    \label{fig:prof1d}
\end{figure}

Error between measured reflectivity and theß mean predicted reflectivity vary across cloud regimes, with predictions of low and thin clouds attaining the lowest RMSE, and larger RMSE in deep clouds (Figure~\ref{fig:rmse_by_regime}). Among the test set (Figure~\ref{fig:prof1d}), CERBERUS displays the most confident and correct predictions of stratiform and low clouds (bottom~row), but consistently struggles to predict complex multilayer clouds (top~left). These results mirror the performance of SatMAE in \citet{girtsou_3d_2025}: nimbostratus and the prevalent deep convective clouds over the SGP are most challenging to predict. By weighting RMSE according to cloud regime prevalence in the European geostationary satellite record \citep{girtsou_3d_2025}, we find equivalent or improved performance (depending on the metric) across cloudy scenes relative to previous 3D cloud predictions (Table~\ref{tab:rmse_table}).

\begin{figure}[!t]
    \centering
    \includegraphics[width=\linewidth]{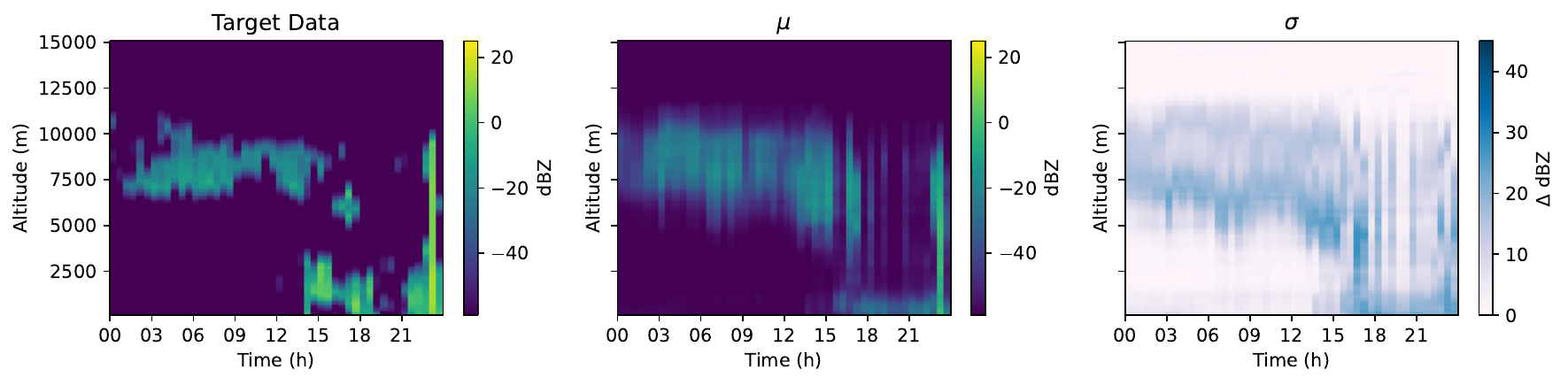}
    \caption{Illustration of the time-evolving reflectivity on Jan 29, 2025 (test set): true KAZR data (far left); and CERBERUS model mean and uncertainty (panels 2-3).}
    \label{fig:jan29}
\end{figure}

Time-resolved composites (Figures~\ref{fig:prof1d},~\ref{fig:feb11},~\ref{fig:feb12}) reveal that CERBERUS captures coherent cloud evolution across the test period, for instance capturing the development of a decoupled precipitating and later convective cloud beneath an initial anvil on Jan 29, 2025. Satellite measurements often saturate in the upper cloud layer, providing limited conditioning on the decoupled clouds underneath. However, CERBERUS does tend to predict a broader reflectivity distribution at these challenging altitudes (model spread in Figure~\ref{fig:prof1d} top left; right panel in Figure~\ref{fig:jan29}), indicating that learned uncertainty reflects physically meaningful ambiguity.  As in \citep{bruning_artificial_2024}, the ZIB mean predictions exhibit overly smooth cloud boundaries, but the distribution spread predicted by CERBERUS add value by indicating these structural uncertainties, with larger spread at and below cloud base and in multilayer clouds, and with less uncertainty in the convective core.

\section{Conclusions \& Future Work}
CERBERUS uses a three-headed encoder-decoder architecture to produce probabilistic estimates of vertically-resolved cloud reflectivity conditioned on 2D satellite fields and near-surface meteorological variables. Despite its simplicity and 1D-profile predictions, CERBERUS produces coherent reflectivity fields with uncertainty estimates that reflect the non-uniqueness of mapping a 2D satellite image to a 3D cloud. Future work will extend this framework toward predictions of model-relevant cloud microphysical quantities such as cloud water content and droplet size distributions, incorporating additional data from both ground-based Doppler radar as well as from global high-resolution models \citep[e.g.]{donahue_exascale_2024}. We anticipate that this extension to global microphysical data will necessitate more expressive architectures such as mixture-density predictions \citep{bishop_mixture_1994} or transformer-based encoding \cite{cong_satmae_2023}, as well as additional conditional inputs like location embedding or broader satellite horizontal context.

\section*{Acknowledgments}
This work was performed under the auspices of the U.S. Department of Energy by Lawrence Livermore National Laboratory (LLNL) under Contract DE-AC52-07NA27344 and supported by the Laboratory Directed Research and Development Program (LDRD), project number 25-ERD-045. The authors have declared that none of them have any competing interests. Released under IM number LLNL-CONF-2015468.

Claude and ChatGPT were used to assist in debugging code, result visualization, and editorial support. All concepts related to model structure, datasets, training strategy, and qualitative analysis were developed by the authors.

All source code, analysis notebooks, and post-processed data used to produce the results in this paper are archived at \url{https://zenodo.org/records/19242435}.

\bibliography{ML_inference}

@article{donahue_exascale_2024,
	title = {To {Exascale} and {Beyond}—{The} {Simple} {Cloud}-{Resolving} {E3SM} {Atmosphere} {Model} ({SCREAM}), a {Performance} {Portable} {Global} {Atmosphere} {Model} for {Cloud}-{Resolving} {Scales}},
	volume = {16},
	copyright = {© 2024 The Author(s). Journal of Advances in Modeling Earth Systems published by Wiley Periodicals LLC on behalf of American Geophysical Union.},
	issn = {1942-2466},
	url = {https://onlinelibrary.wiley.com/doi/abs/10.1029/2024MS004314},
	doi = {10.1029/2024MS004314},
	language = {en},
	number = {7},
	urldate = {2024-12-16},
	journal = {Journal of Advances in Modeling Earth Systems},
	author = {Donahue, A. S. and Caldwell, P. M. and Bertagna, L. and Beydoun, H. and Bogenschutz, P. A. and Bradley, A. M. and Clevenger, T. C. and Foucar, J. and Golaz, C. and Guba, O. and Hannah, W. and Hillman, B. R. and Johnson, J. N. and Keen, N. and Lin, W. and Singh, B. and Sreepathi, S. and Taylor, M. A. and Tian, J. and Terai, C. R. and Ullrich, P. A. and Yuan, X. and Zhang, Y.},
	year = {2024},
	note = {\_eprint: https://onlinelibrary.wiley.com/doi/pdf/10.1029/2024MS004314},
	pages = {e2024MS004314},
}

@article{morrison_confronting_2020,
	title = {Confronting the {Challenge} of {Modeling} {Cloud} and {Precipitation} {Microphysics}},
	volume = {12},
	issn = {1942-2466},
	doi = {10.1029/2019MS001689},
	language = {en},
	number = {8},
	journal = {Journal of Advances in Modeling Earth Systems},
	author = {Morrison, Hugh and Lier‐Walqui, Marcus van and Fridlind, Ann M. and Grabowski, Wojciech W. and Harrington, Jerry Y. and Hoose, Corinna and Korolev, Alexei and Kumjian, Matthew R. and Milbrandt, Jason A. and Pawlowska, Hanna and Posselt, Derek J. and Prat, Olivier P. and Reimel, Karly J. and Shima, Shin-Ichiro and Diedenhoven, Bastiaan van and Xue, Lulin},
	year = {2020},
}

@misc{bishop_mixture_1994,
	address = {Birmingham},
	type = {Monograph},
	title = {Mixture density networks},
	copyright = {cc\_by\_nc\_nd\_4},
	url = {https://publications.aston.ac.uk/id/eprint/373/},
	language = {en-GB},
	urldate = {2026-01-27},
	publisher = {Aston University},
	author = {Bishop, Christopher M.},
	year = {1994},
	note = {Num Pages: 438543},
}

@book{boucher_clouds_2013,
	address = {Cambridge, UK and New York, NY, USA},
	title = {Clouds and {Aerosols}. {In}: {Climate} {Change} 2013: {The} {Physical} {Science} {Basis}. {Contribution} of {Working} {Group} {I} to the {Fifth} {Assessment} {Report} of the {Intergovernmental} {Panel} on {Climate} {Change}},
	publisher = {Cambridge University Press},
	author = {Boucher, Olivier and Randall, D. and Bretherton, C. and Feingold, Graham and Forster, P. and Kerminen, V.-M. and Kondo, Y. and Liao, H. and Lohmann, U. and Rasch, Philip J and Satheesh, S.K. and Sherwood, S. and Stevens, B. and Zhang, X.Y.},
	year = {2013},
}

@article{lamb_perspectives_2026,
	title = {Perspectives on {Systematic} {Cloud} {Microphysics} {Scheme} {Development} {With} {Machine} {Learning}},
	volume = {18},
	issn = {1942-2466},
	url = {https://onlinelibrary.wiley.com/doi/abs/10.1029/2025MS005341},
	doi = {10.1029/2025MS005341},
	language = {en},
	number = {1},
	urldate = {2026-01-27},
	journal = {Journal of Advances in Modeling Earth Systems},
	author = {Lamb, Kara D. and Singer, Clare E. and Loftus, Kaitlyn and Morrison, Hugh and Powell, Margaret and Ko, Joseph and Buch, Jatan and Hu, Arthur Z. and van Lier Walqui, Marcus and Gentine, Pierre},
	year = {2026},
	note = {\_eprint: https://agupubs.onlinelibrary.wiley.com/doi/pdf/10.1029/2025MS005341},
	pages = {e2025MS005341},
}

@article{savitzky_smoothing_1964,
	title = {Smoothing and {Differentiation} of {Data} by {Simplified} {Least} {Squares} {Procedures}.},
	volume = {36},
	issn = {0003-2700},
	url = {https://doi.org/10.1021/ac60214a047},
	doi = {10.1021/ac60214a047},
	number = {8},
	urldate = {2026-01-18},
	journal = {Analytical Chemistry},
	publisher = {American Chemical Society},
	author = {Savitzky, Abraham. and Golay, M. J. E.},
	month = jul,
	year = {1964},
	pages = {1627--1639},
}

@misc{KAZR_SGP,
  title = {{ARSCLKAZRBND1KOLLIAS}, {SGP}, 2020-2024},
  howpublished = {https://adc.arm.gov/discovery/results},
  note = {Accessed: 2025-04-22},
  shorthand = {KAZR}
}

@misc{GOES_SGP,
  title = {{VISSTGRIDG16V4MINNIS}, {SGP}, 2020-2024},
  howpublished = {https://adc.arm.gov/discovery/results},
  note = {Accessed: 2025-04-24},
  shorthand = {G16},
}

@misc{MERRA2,
  title = {{M2I1NXASM.5.12.4}:inst1\_2d\_asm\_{N}x},
  howpublished = {https://disc.gsfc.nasa.gov/datasets?project=MERRA-2},
  note = {Accessed: 2025-04-22},
  shorthand = {MERRA2}
}

@article{leinonen_reconstruction_2019,
	title = {Reconstruction of {Cloud} {Vertical} {Structure} {With} a {Generative} {Adversarial} {Network}},
	volume = {46},
	copyright = {©2019. American Geophysical Union. All Rights Reserved.},
	issn = {1944-8007},
	url = {https://onlinelibrary.wiley.com/doi/abs/10.1029/2019GL082532},
	doi = {10.1029/2019GL082532},
	language = {en},
	number = {12},
	urldate = {2024-10-04},
	journal = {Geophysical Research Letters},
	author = {Leinonen, Jussi and Guillaume, Alexandre and Yuan, Tianle},
	year = {2019},
	note = {\_eprint: https://onlinelibrary.wiley.com/doi/pdf/10.1029/2019GL082532},
	pages = {7035--7044},
}

@article{bruning_artificial_2024,
	title = {Artificial intelligence ({AI})-derived {3D} cloud tomography from geostationary {2D} satellite data},
	volume = {17},
	issn = {1867-1381},
	url = {https://amt.copernicus.org/articles/17/961/2024/},
	doi = {10.5194/amt-17-961-2024},
	language = {English},
	number = {3},
	urldate = {2024-10-07},
	journal = {Atmospheric Measurement Techniques},
	author = {Brüning, Sarah and Niebler, Stefan and Tost, Holger},
	month = feb,
	year = {2024},
	note = {Publisher: Copernicus GmbH},
	pages = {961--978},
}

@misc{ermis_global_2025,
	title = {Global {3D} {Reconstruction} of {Clouds} \& {Tropical} {Cyclones}},
	url = {http://arxiv.org/abs/2511.04773},
	doi = {10.48550/arXiv.2511.04773},
	urldate = {2025-12-09},
	publisher = {arXiv},
	author = {Ermis, Shirin and Aybar, Cesar and Freischem, Lilli and Girtsou, Stella and Bintsi, Kyriaki-Margarita and Salas-Porras, Emiliano Diaz and Eisinger, Michael and Jones, William and Jungbluth, Anna and Tremblay, Benoit},
	month = nov,
	year = {2025},
	note = {arXiv:2511.04773 [cs]},
}

@misc{girtsou_3d_2025,
	title = {{3D} {Cloud} reconstruction through geospatially-aware {Masked} {Autoencoders}},
	url = {http://arxiv.org/abs/2501.02035},
	doi = {10.48550/arXiv.2501.02035},
	urldate = {2025-12-10},
	publisher = {arXiv},
	author = {Girtsou, Stella and Salas-Porras, Emiliano Diaz and Freischem, Lilli and Massant, Joppe and Bintsi, Kyriaki-Margarita and Castiglione, Guiseppe and Jones, William and Eisinger, Michael and Johnson, Emmanuel and Jungbluth, Anna},
	month = jan,
	year = {2025},
	note = {arXiv:2501.02035 [cs]},
}

@misc{cong_satmae_2023,
	title = {{SatMAE}: {Pre}-training {Transformers} for {Temporal} and {Multi}-{Spectral} {Satellite} {Imagery}},
	shorttitle = {{SatMAE}},
	url = {http://arxiv.org/abs/2207.08051},
	doi = {10.48550/arXiv.2207.08051},
	urldate = {2025-12-12},
	publisher = {arXiv},
	author = {Cong, Yezhen and Khanna, Samar and Meng, Chenlin and Liu, Patrick and Rozi, Erik and He, Yutong and Burke, Marshall and Lobell, David B. and Ermon, Stefano},
	month = jan,
	year = {2023},
	note = {arXiv:2207.08051 [cs]},
}

@article{kollias_millimeter-wavelength_2007,
	chapter = {Bulletin of the American Meteorological Society},
	title = {Millimeter-{Wavelength} {Radars}: {New} {Frontier} in {Atmospheric} {Cloud} and {Precipitation} {Research}},
	shorttitle = {Millimeter-{Wavelength} {Radars}},
	url = {https://journals.ametsoc.org/view/journals/bams/88/10/bams-88-10-1608.xml},
	doi = {10.1175/BAMS-88-10-1608},
	language = {en},
	urldate = {2025-05-14},
	author = {Kollias, P. and Clothiaux, E. E. and Miller, M. A. and Albrecht, B. A. and Stephens, G. L. and Ackerman, T. P.},
	month = oct,
	year = {2007},
}

@misc{perez_film_2017,
	title = {{FiLM}: {Visual} {Reasoning} with a {General} {Conditioning} {Layer}},
	shorttitle = {{FiLM}},
	url = {http://arxiv.org/abs/1709.07871},
	doi = {10.48550/arXiv.1709.07871},
	urldate = {2026-03-11},
	publisher = {arXiv},
	author = {Perez, Ethan and Strub, Florian and Vries, Harm de and Dumoulin, Vincent and Courville, Aaron},
	month = dec,
	year = {2017},
	note = {arXiv:1709.07871 [cs]},
}

@article{gneiting_strictly_2007,
	title = {Strictly {Proper} {Scoring} {Rules}, {Prediction}, and {Estimation}},
	volume = {102},
	issn = {0162-1459},
	url = {https://doi.org/10.1198/016214506000001437},
	doi = {10.1198/016214506000001437},
	number = {477},
	urldate = {2026-03-24},
	journal = {Journal of the American Statistical Association},
	publisher = {Taylor \& Francis},
	author = {Gneiting, Tilmann and Raftery, Adrian E},
	month = mar,
	year = {2007},
	note = {\_eprint: https://doi.org/10.1198/016214506000001437},
	pages = {359--378},
}

@misc{akiba_optuna_2019,
	title = {Optuna: {A} {Next}-generation {Hyperparameter} {Optimization} {Framework}},
	shorttitle = {Optuna},
	url = {http://arxiv.org/abs/1907.10902},
	doi = {10.48550/arXiv.1907.10902},
	urldate = {2026-03-26},
	publisher = {arXiv},
	author = {Akiba, Takuya and Sano, Shotaro and Yanase, Toshihiko and Ohta, Takeru and Koyama, Masanori},
	month = jul,
	year = {2019},
	note = {arXiv:1907.10902 [cs]},
}
\bibliographystyle{iclr2026_conference}

\newpage
\appendix
\renewcommand{\thefigure}{A.\arabic{figure}}
\renewcommand{\thetable}{A.\arabic{table}}
\setcounter{figure}{0}
\setcounter{table}{0}
\section{Appendix}

\subsection{Data Filtering Criteria}
\label{sec:Data}
\begin{figure}[H]
    \centering
    \includegraphics[width=0.4\linewidth]{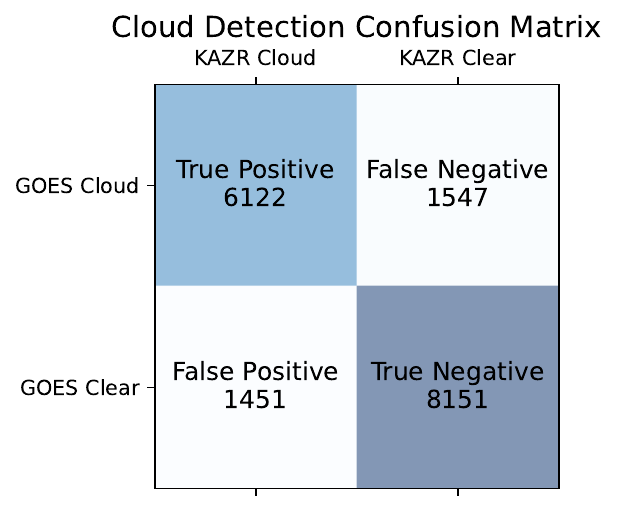}
    \caption{Clear-sky vs cloud detection from GOES and KAZR at the ARM SGP site in 2022, aggregated across altitudes. Only the "True Positive" data are included in the training and validation datasets.}
    \label{fig:confusion}
\end{figure}

The follow criteria are applied to remove radar-radiometer pairs which are not cloudy, or are not consistent between the measuring instruments (KAZR, GOES). All 4 criteria are applied to the training and validation datasets, but the 4th criterion is not applied to data in the test set in order to mimic the case where a ground-truth reflectivity measurement is not available for comparison.
\begin{enumerate}
    \item $\max_z Z(z) > -40$dBZ, where $Z$ is KAZR reflectivity (in dBZ) and $z$ is altitude; removes artifacts.
    \item Cloud thickness (from KAZR) $> 200m$, removes very thin clouds.
    \item Both KAZR and GOES have valid (non-NaN) measurements at the SGP site (see Figure \ref{fig:confusion}).
    \item $|CTH_{GOES} - CTH_{KAZR}| < \sigma(|CTH_{GOES} - CTH_{KAZR}|)$ where $CTH$ is cloud-top height and $\sigma$ is the standard deviation across the training/validation dataset; removes inconsistent scenes where GOES and KAZR may not be measuring the same cloud.
\end{enumerate}

\begin{figure}[H]
    \centering
    \includegraphics[width=0.4\linewidth]{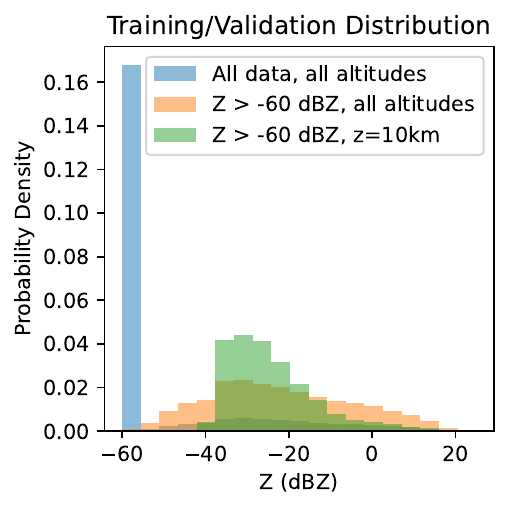}
    \caption{Distribution of aggregated target reflectivities from the training and validation KAZR data, aggregated across altitude (blue, orange) and timestamps (all). Blue includes all training/validation data with NaN-reflectivities filled with the -60dBZ detection threshold; orange indicates the PDF with these non-cloudy ($\leq -60$~dBZ) data removed; and green indicates the PDF of all reflectivities at 10km altitude aggregated across training/validation.}
    \label{fig:pdf}
\end{figure}

%%%%%%%%%%%%%%%%%%%%%%%%%%%%%%%%%%%%%%%%%%%%%%%%%%%%%%%%%%%%%%%%%%%%%%%%%
\subsection{Additional Metrics}
\label{sec:metrics}
\begin{figure}[H]
    \centering
    \includegraphics[width=\linewidth]{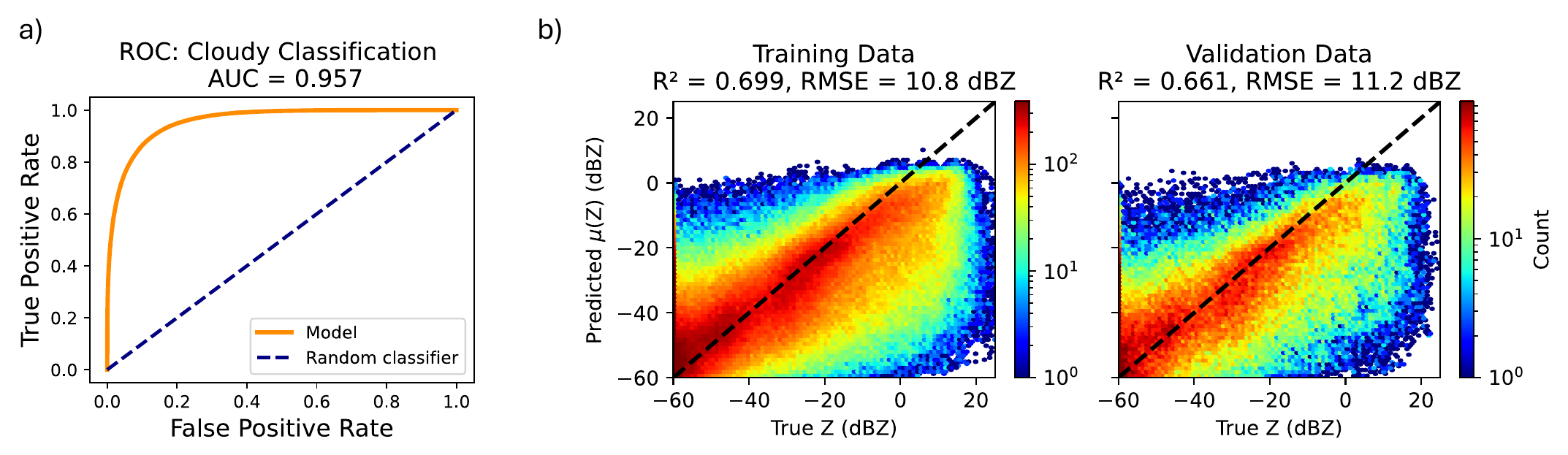}
    \caption{(a) Receiver operating characteristic (ROC) curve for the classification head $\pi$ of CERBERUS evaluated on validation data; (b) parity between the true reflectivity values and the mean of the ZIB prediction, aggregated across altitudes and timestamps in the training and validation sets, respectively.}
    \label{fig:rsq}
\end{figure}

In addition to the negative log-likelihood (NLL) used to train CERBERUS (Equation~1) we report additional traditional metrics including the Receiver Operating Curve (ROC) and Area Under the Curve (AUC), the correlation coefficient $R^2$, and the root-mean squared error (RMSE) between individual reflectivity predictions per altitude (Figure~\ref{fig:rsq}). Specifically, RMSE is evaluated as the difference between the measurement $y$ and the mean of the distribution prediction.

In addition to these traditional metrics, we further consider the Continuous Ranked Probability Score (CRPS) and the Energy of our model \citep{gneiting_strictly_2007}, which are scalar and vector (respectively) scoring functions that treat the difference between a distribution and an observation. Specifically, the CRPS measures the elementwise difference between the predicted CDF $F$ and empirical CDF corresponding to the measurement $y$:
\begin{align}
    \texttt{CRPS}(F,y) &= \int \big( F(x) - H(x-y)\big)^2 dx \\
    &= \E_{X \sim F} |X - y| - \frac{1}{2}\E_{X,X' \sim F}|X-X'|,
\end{align}
where $H$ denotes the Heaviside function and $\mathop{\mathbb{E}}$ denotes expectation. For our beta-type distributions, we take 100 samples from the predicted CDF to compute the empirical CRPS. CRPS reduces to mean absolute error in the case of a deterministic prediction.

Whereas CRPS is computed independently for each altitude of a predicted cloud profile, the Energy is a score on a multivariate prediction using vector norms:
\begin{align}
    \texttt{Energy}(F,y) &= \E_{X \sim F} ||X - y|| - \frac{1}{2}\E_{X,X' \sim F}||X-X'||.
\end{align}
We adopt the 2-norm convention for this work. In that case, the Energy is proportional to the root-mean squared error in the case of a deterministic prediction, differing by a factor of the square root of the vector dimension ($\sqrt{128}$ in our case). In Figures~\ref{fig:prof1d} and \ref{fig:baselines} and Table~\ref{tab:rmse_table}, Energy is reported with this vector-length scaling taken into account in order to make it directly comparable to RMSE.

%%%%%%%%%%%%%%%%%%%%%%%%%%%%%%%%%%%%%%%%%%%%%%%%%%%%%%%%%%%%%%%%%%%%%%%%

\subsection{Comparison to Non-Zero-Inflated Baselines}
\label{sec:baselines}
To verify improvement in model skill from the probabilistic ZIB prediction target of CERBERUS, we compare its performance to two additional baseline models. The first is a purely deterministic model, which uses a single-headed output layer to predict the reflectivity and is trained using mean-squared error as the loss function. The second is a beta distribution target \textit{without} a ZI component: this two headed model predicts only the $\alpha$ and $\beta$ parameters of the beta distributed reflectivity using two prediction heads, and is trained using the negative log likelihood with the standard beta distribution. All three cases (deterministic, beta, and ZIB) utilize the same model structure otherwise and differ only in the number of output heads.

\begin{figure}[H]
    \centering
    \includegraphics[width=0.7\linewidth]{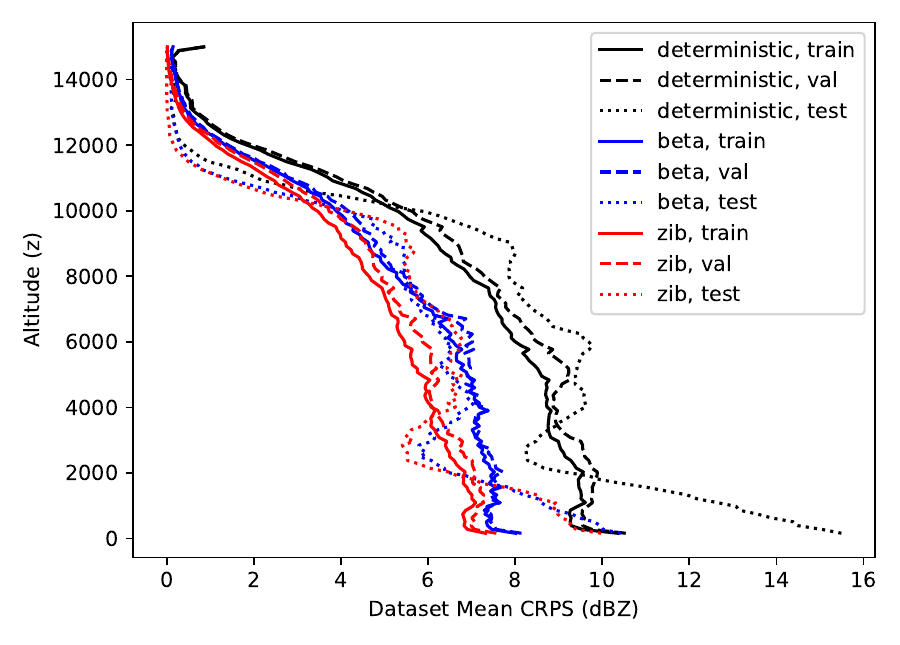}
    \caption{Continuously ranked probability score (or MAE, for deterministic case) for the ZIB CERBERUS model and the non-ZI beta and deterministic baselines. CRPS is computed independently at each altitude for a given reflectivity measurement-prediction pair, and we report the mean CRPS across the training, validation, and testing datasets, respectively.}
    \label{fig:crps}
\end{figure}

\begin{figure}[H]
    \centering
    \includegraphics[width=\linewidth]{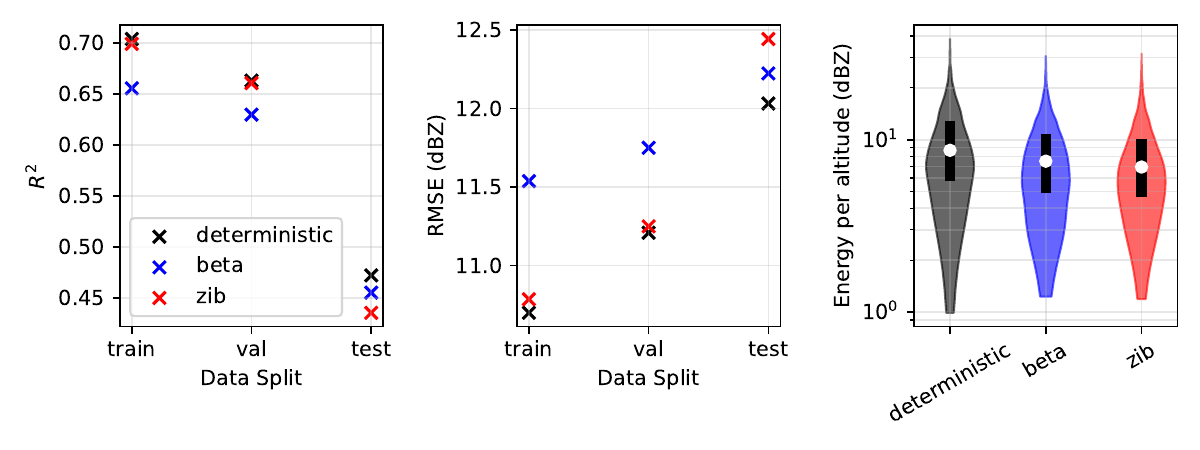}
    \caption{Correlation coefficient $R^2$ (left), RMSE (middle), and scaled Energy (RMSE) for the ZIB model and baselines. The first two metrics are reported for training, validation, and testing data splits. The Energy / RMSE is reported only for the validation dataset; Energy is scaled by the vector size ($\times 1/\sqrt{128}$) to make it directly comparable to the RMSE that is reported for the deterministic model.}
    \label{fig:baselines}
\end{figure}

Analysis of the CRPS (or mean-absolute error in the deterministic case) in Figure~\ref{fig:crps} reveals that indeed, the probabilistic predictions from the ZIB and beta models outperforms the deterministic model across altitudes. Furthermore, the ZIB predictions add value on the order of a 1dBZ reduction in error below 8~km altitude relative to a non-ZI counterpart. Aggregated results reported in Figure~\ref{fig:baselines} tell a similar story: in deterministic metrics such as RMSE and correlation coefficient, the expectation value of the ZIB model consistently outperforms that of the 2-headed beta model, achieving comparable performance to the deterministic case. In the probabilistic Energy metric (RMSE for deterministic case), the ZIB model consistently displays the smallest distance from measurements. These results validate and motivate the choice to use a zero-inflated and probabilistic prediction target over deterministic and non-ZI alternatives for this particular reflectivity application.

%%%%%%%%%%%%%%%%%%%%%%%%%%%%%%%%%%%%%%%%%%%%%%%%%%%%%%%%%%%%%%%%%%%%%%%%%
\newpage
\subsection{Training Characteristics}
\label{sec:training}
\begin{figure}[H]
    \centering
    \includegraphics[width=0.5\linewidth]{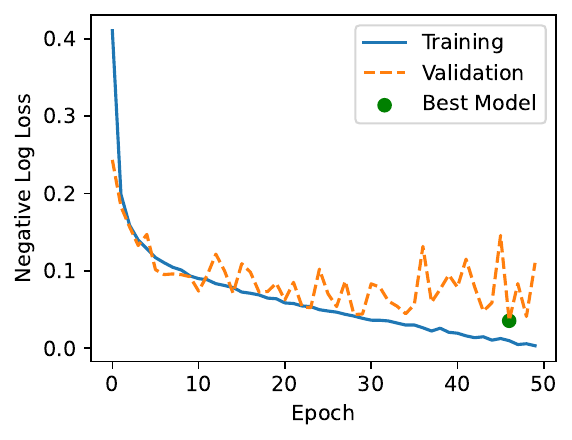}
    \caption{Training and validation loss for the presented ZIB configuration of CERBERUS.}
    \label{fig:loss}
\end{figure}

%%%%%%%%%%%%%%%%%%%%%%%%%%%%%%%%%%%%%%%%%%%%%%%%%%%%%%%%%%%%%%%%%%%%%%%%%
\newpage
\subsection{Performance by Cloud Regime}
\label{sec:cr}
Table~\ref{tab:rmse_table} reports both RMSE (between data and predicted mean) and the scaled Energy score for CERBERUS. The results reported for SatMAE are taken from \cite{girtsou_3d_2025}. Cloud regimes in the SGP dataset are determined according to cloud-top-height thresholds (2km, 6km) and maximum cloud reflectivity thresholds (-20dBZ, 0dBZ) from the KAZR satellite to permit inclusion of nighttime clouds that would be excluded from an optical depth threshold.
\begin{table}[H]
    \centering
    \begin{tabular}{c|c|c}
        Cloud Regime & SatMAE RMSE in dBZ & CERBERUS RMSE / Energy in dBZ \\
        (Abbreviation) & (SEVIRI prevalence) & (CERBERUS prevalence) \\
        \hline
        \hline
        Cumulus (Cu) & 7.0 (0.8\%) & 3.4 / 2.9 (7.8\%) \\
        Stratocumulus (Sc) & 7.5 (1.4\%) & 4.7 / 3.9 (4.1\%) \\
        Stratus (St) & 4.8 (3.0\%) & 6.2 / 5.1 (0.1\%) \\
        Altocumulus (AC) & 8.4 (0.6\%) & 5.6 / 4.7 (9.6\%) \\
        Altostratus (AS) & 9.8 (0.4\%) & 8.1 / 6.5 (10.1\%) \\
        Nimbostratus (NS) & 12.5 (0.0\%) & 14.5 / 11.4 (3.4\%) \\
        Cirrus (Ci) & 4.8 (0.6\%) & 7.1 / 5.8 (15.0\%) \\
        Cirrostratus (CS) & N/A & 10.8 / 8.6 (31.3\%) \\
        Deep Convection (DC) & 10.3 (0.6\%) & 16.2 / 12.8 (18.5\%) \\
        \hline
        Native-mean (non-clear sky) & 6.6 & 9.7 / 7.8 \\
        \hline
        SEVIRI-weighted mean & 6.6 & 6.5 / 5.3 \\
        \hline
    \end{tabular}
    \caption{RMSE (mean across validation set samples), scaled Energy (for CEBERUS), and frequency of various cloud regimes, including results from CERBERUS at the SGP and results reported in \citet{girtsou_3d_2025} for the SatMAE model over the MSG/SEVIRI satellite range. Clear-sky scenes are excluded from this analysis. Overall RMSE is reported across cloud regimes according to raw data weighting and according to SEVIRI cloud regime weighting.  These data correspond to Figure~\ref{fig:rmse_by_regime}.}
    \label{tab:rmse_table}
\end{table}
%%%%%%%%%%%%%%%%%%%%%%%%%%%%%%%%%%%%%%%%%%%%%%%%%%%%%%%%%%%%%%%%%%%%%%%%%
\newpage
\subsection{Feature Importance}
\label{sec:pfi}
\begin{figure}[h!]
    \centering
    \includegraphics[width=0.7\linewidth]{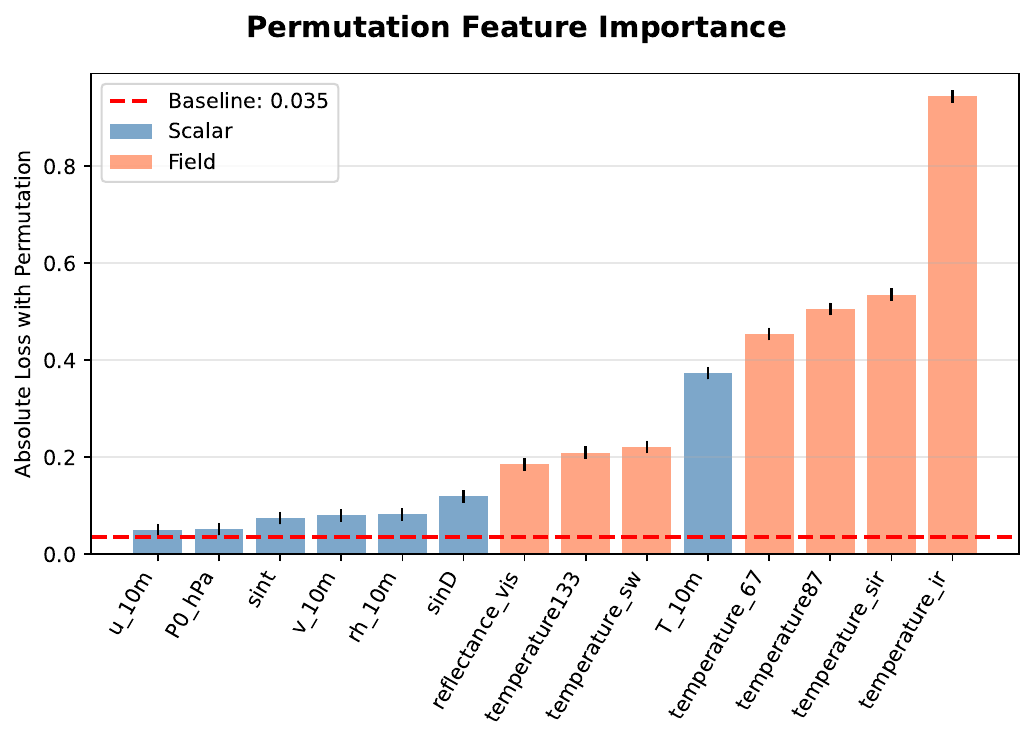}
    \caption{Permutation feature importance with the original model loss (negative log likelihood) plotted as a red-dashed line. Features are plotting in order of least important/smallest change to loss, to most important, and are colored according to field type (GOES) or scalar type (meteorology/reanalysis).}
    \label{fig:pfi}
\end{figure}

%%%%%%%%%%%%%%%%%%%%%%%%%%%%%%%%%%%%%%%%%%%%%%%%%%%%%%%%%%%%%%%%%%%
\newpage
\section{Additional Time-height Reflectivity Demonstrations}
\label{sec:timeheight}
\begin{figure}[h!]
    \centering
    \includegraphics[width=\linewidth]{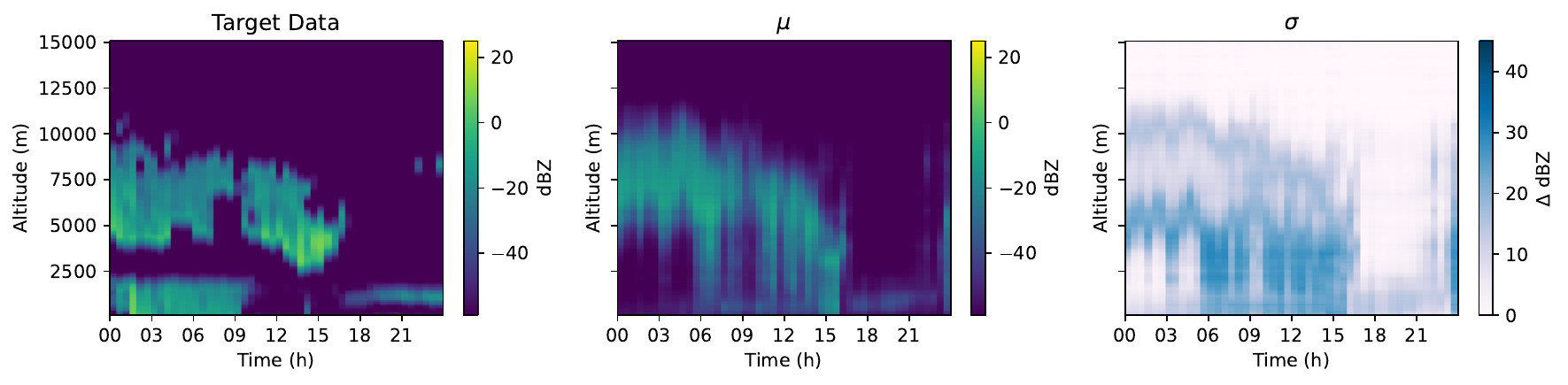}
    \caption{As in Figure~\ref{fig:jan29}, for February 11, 2025 (test set).}
    \label{fig:feb11}
\end{figure}

\begin{figure}[h!]
    \centering
    \includegraphics[width=\linewidth]{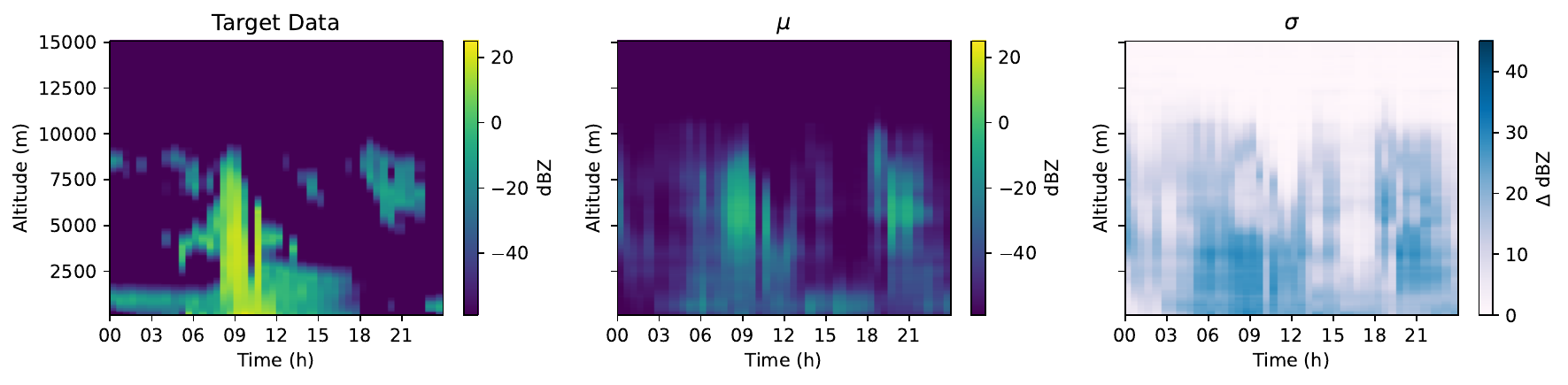}
    \caption{As in Figure~\ref{fig:jan29}, for February 12, 2025 (test set).}
    \label{fig:feb12}
\end{figure}

\end{document}